\documentclass[runningheads]{llncs}
\usepackage{graphicx}
\usepackage{xcolor}
\usepackage[T1]{fontenc}
\usepackage{multirow}
\usepackage{setspace}
\usepackage{booktabs}
\usepackage{stackengine}
\usepackage{soul}
\usepackage{fontenc}

\begin{document}

\title{Spatially Dependent U-Nets: Highly Accurate Architectures for Medical Imaging Segmentation}
\authorrunning{J. B. S. Carvalho et al.}
\titlerunning{SDU-Nets: Highly Accurate Architectures for MI Segmentation}

% \author{Anonymous}
% \institute{Anonymous Organization\\
% \email{***@*******.***}}
\author{Jo\~ao B. S. Carvalho\inst{1} \and
Jo\~ao A. Santinha\inst{2,3}  \and
\DJ or\dj e Miladinovi\'{c}\inst{1} \and \\
Joachim M. Buhmann\inst{1} }

\institute{Institute for Machine Learning, ETH Zurich, Switzerland \and
Instituto Superior T\'ecnico, Universidade de Lisboa, Portugal\and
Computational Clinical Imaging Group, Champalimaud Foundation, Portugal}
% \author{First Author\inst{1}\orcidID{0000-1111-2222-3333} \and
% Second Author\inst{2,3}\orcidID{1111-2222-3333-4444} \and
% Third Author\inst{3}\orcidID{2222--3333-4444-5555}}
% %
% \authorrunning{F. Author et al.}
% First names are abbreviated in the running head.
% If there are more than two authors, 'et al.' is used.

% \institute{Princeton University, Princeton NJ 08544, USA \and
% Springer Heidelberg, Tiergartenstr. 17, 69121 Heidelberg, Germany
% \email{lncs@springer.com}\\
% \url{http://www.springer.com/gp/computer-science/lncs} \and
% ABC Institute, Rupert-Karls-University Heidelberg, Heidelberg, Germany\\
% \email{\{abc,lncs\}@uni-heidelberg.de}}
%
\maketitle              % typeset the header of the contribution

\begin{abstract} 
 In clinical practice, regions of interest in medical imaging often need to be identified through a process of precise image segmentation. 
 The quality of this image segmentation step critically affects the subsequent clinical assessment of the patient status.
 To enable high-accuracy, automatic image segmentation, we introduce a novel deep neural network architecture that exploits the inherent spatial coherence of anatomical structures and is well-equipped to capture long-range spatial dependencies in the segmented pixel/voxel space.
 In contrast to the state-of-the-art solutions based on convolutional layers, our approach leverages on recently introduced spatial dependency layers that have an unbounded receptive field and explicitly model the inductive bias of spatial coherence.
 Our method performs favourably to commonly used U-Net and U-Net++ architectures as demonstrated by improved Dice and Jaccard score in three different medical segmentation tasks:  nuclei segmentation in microscopy images, polyp segmentation in colonoscopy videos, and liver segmentation in abdominal CT scans.
 \keywords{Spatial Dependency Networks; Medical Image Segmentation, U-Net.}
\end{abstract}%

\section{Introduction}
Medical Imaging (MI) enables the in-vivo and non-invasive visualization of structural, molecular, and functional information inside the human body \cite{xue2018segan}. Imaging the internal tissues of a patient facilitates the diagnosis, prognosis, and treatment planning \cite{WANG20161262,gonzalez2015role,son2021diagnostic}. MI-driven patient assessment is normally based on the (subjective) opinion of a physician but can be automated e.g. using recent developments in artificial intelligence \cite{ronneberger2015unet} with the potential of accelerating patient management and removing the subjectivity from clinical decision-making \cite{GIGER2018512}.
Whether the interpretation of MI data is done by doctors or automatic methods, a typical preceding step in the clinical pipeline common to both is to perform image segmentation to reduce the dimensionality of data and highlight regions of interest.
Since manual segmentation is a time-consuming and tedious task for physicians, many algorithms have been developed to automate this process. Arguably most successful approaches are based on deep learning \cite{Goodfellow-et-al-2016}, utilizing fully convolutional networks (FCN) and encoder-decoder-based convolutional networks such as the U-net \cite{ronneberger2015unet}. Extrapolating on their remarkable progress over the past few years \cite{zhou2019unet++,isensee2018no}, we speculate that neural network-based algorithms are very likely to match human-level performance in the near future. The hereby proposed approach is a step in that direction.

\begin{figure}[t]
\includegraphics[width=\textwidth]{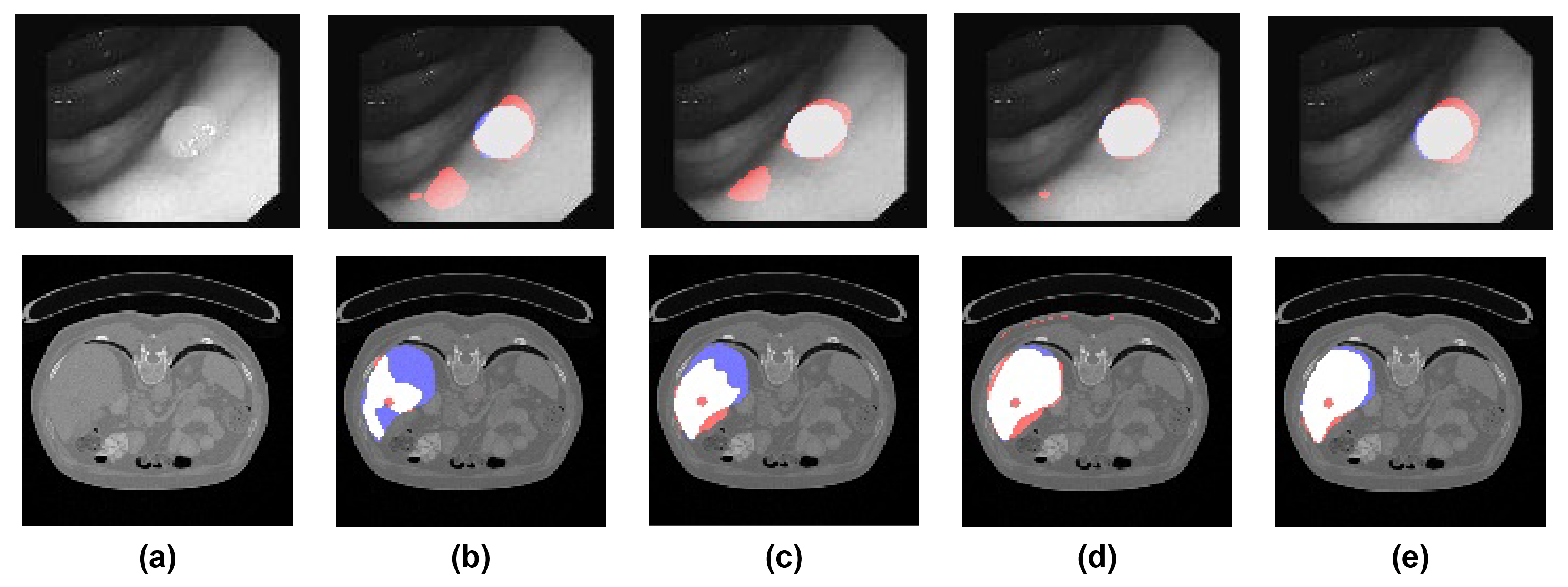}
\caption{\textbf{Qualitative analysis of image segmentation for different competing methods (best viewed in color).}  \textbf{(a)} original image; \textbf{(b)} U-Net \cite{ronneberger2015unet}; \textbf{(c)} U-Net++ \cite{zhou2019unet++}; \textbf{(d)} SDU-Net (ours); \textbf{(e)} SDNU-Net (ours);
\textbf{(top row)} \emph{polyp segmentation}.
The baseline U-Net architectures mistakenly identify a region in the bottom left corner as relevant due to the local features.
However, when observing the image as a whole, it is clear that there is a single region of interest, as marked successfully by our networks.
\textbf{(bottom row)} \emph{liver segmentation}.
In contrast to SDN-based networks, the baseline U-Net architectures are unable to coherently identify the well-shaped region of interest. 
\textbf{(red color)} false positives; \textbf{(blue color)} false negatives;
\textbf{(white color)} correctly predicted pixels.
} \label{figqualitative_comparison}
\end{figure}

The performance of automatic medical image segmentation critically depends on the capability of machine learning algorithms to: \emph{(i)} accurately identify intensity discontinuities or edges as object boundaries that characterize local textures; and \emph{(ii)} account for global, contextual information when assessing the relevance of different image regions e.g. understanding image-specific semantics and texture.
Convolutional neural networks excel at \emph{(i)} but often fail at \emph{(ii)} as examplified in Figure \ref{figqualitative_comparison}.
We argue that the main reason is the intrinsically limited receptive field of convolutions coupled with the 'shallow' architecture design typical in medical domain.
This issue is common to practically all state-of-the-art architectures that are based on the U-Net-like design \cite{ronneberger2015unet}.
U-Net improves upon the vanilla encoder-decoder design by enabling fusion of textural features across distinct semantic levels; done by introducing skip connections between encoding and decoding paths at different levels and increasing the number of feature channels in the expansive path. 
However, none of the existing networks in the family of U-Nets is well-equipped to perform non-local similarity comparisons and provide spatially coherent, holistic image segmentation -- one of the key characteristics of human-based medical image segmentation.

To enable U-Net based architectures to produce globally coherent image segmentation, capturing both long and short-range dependencies in the pixel/voxel space, we propose to endow the original architecture with the recently introduced spatial dependency networks (SDNs)\cite{miladinovic2021spatial}.
In particular, we utilize SDN to construct two new variants of U-Nets: \emph{(i)} spatially dependent nested U-Net (SDNU-Net) -- to acquire new state-of-the-art performance in medical image segmentation; and \emph{(ii)} spatially dependent U-Net (SDU-Net) -- a faster and less memory-consuming version of \emph{(i)}.
SDNs have been originally developed in the context of generative image modeling resulting in new state-of-the-art variational autoencoder \cite{miladinovic2021spatial};
in the same spirit, our approach is motivated by the spatially coherent (smooth) nature of anatomical images, where tissue properties are shared homogeneously through neighbouring, identical regions. These structures may additionally carry over long distance dependencies, where feature descriptions are shared with non-adjacent regions. Looking specifically at segmentation tasks, pixel/voxel labelling in medical images not only relies on local textures, but also on features from dissimilar tissues, as seen when medical doctors require a comparative assessment of different image patches to accomplish diagnosis.
We argue that the introduced spatial dependency layers allow U-Net based architectures to better exploit the topological structure in medical images, increasing local coherence of the feature map at the same semantic level, while enhancing how long-distance interactions between features are modeled. Due to widespread use of the CNN schematic in segmentation architectures and the simple integration of spatial dependency layers, replacing common convolutional network building blocks, our implementation is generalizable to other architectures in most medical image segmentation tasks.

We experimentally demonstrate the performance improvements in the quality of image segmentation on three different medical imaging segmentation tasks: \emph{(i)} cell structure segmentation \cite{caicedo2019nucleus}; \emph{(iii)} anatomical structure segmentation \cite{lits_dataset}; and \emph{(iii)} abnormal tissue segmentation \cite{bernal2017comparative}, comprising three distinctive medical imaging methods (respectively, microscopic imaging, colonoscopy video, abdominal CT-scan).
The key contributions of this paper can be summarized as follows:
\subsection*{Key contributions}
\begin{itemize}
    \item Motivated by the necessity of explicit modeling of spatial coherence and non-local dependencies in medical image segmentation, we propose two novel architectures belonging to the family of U-Nets: SDU-Net and SDNU-Net. The key component of our networks is the recently proposed spatial dependency network (SDN) \cite{miladinovic2021spatial}.
    \item We experimentally confirm the superiority of the proposed architectures by comparing them to two widely used baseline U-Net architectures, on three different tasks. Our model evaluation considers both stability of implementation through averaging results across multiple runs, and assessing contributions to performance through a series of ablations studies. 
    \item The proposed architectures' implementation and experiments are disclosed in the frameworks of \textit{Pytorch} \cite{pytorch} and \textit{PytorchLightning} \cite{falcon2019pytorch}.
\end{itemize}

\section{SDU-Net and SDNU-Net Architecture}
\begin{figure}[ht]
\vspace{-0.5cm}
\includegraphics[width=\textwidth]{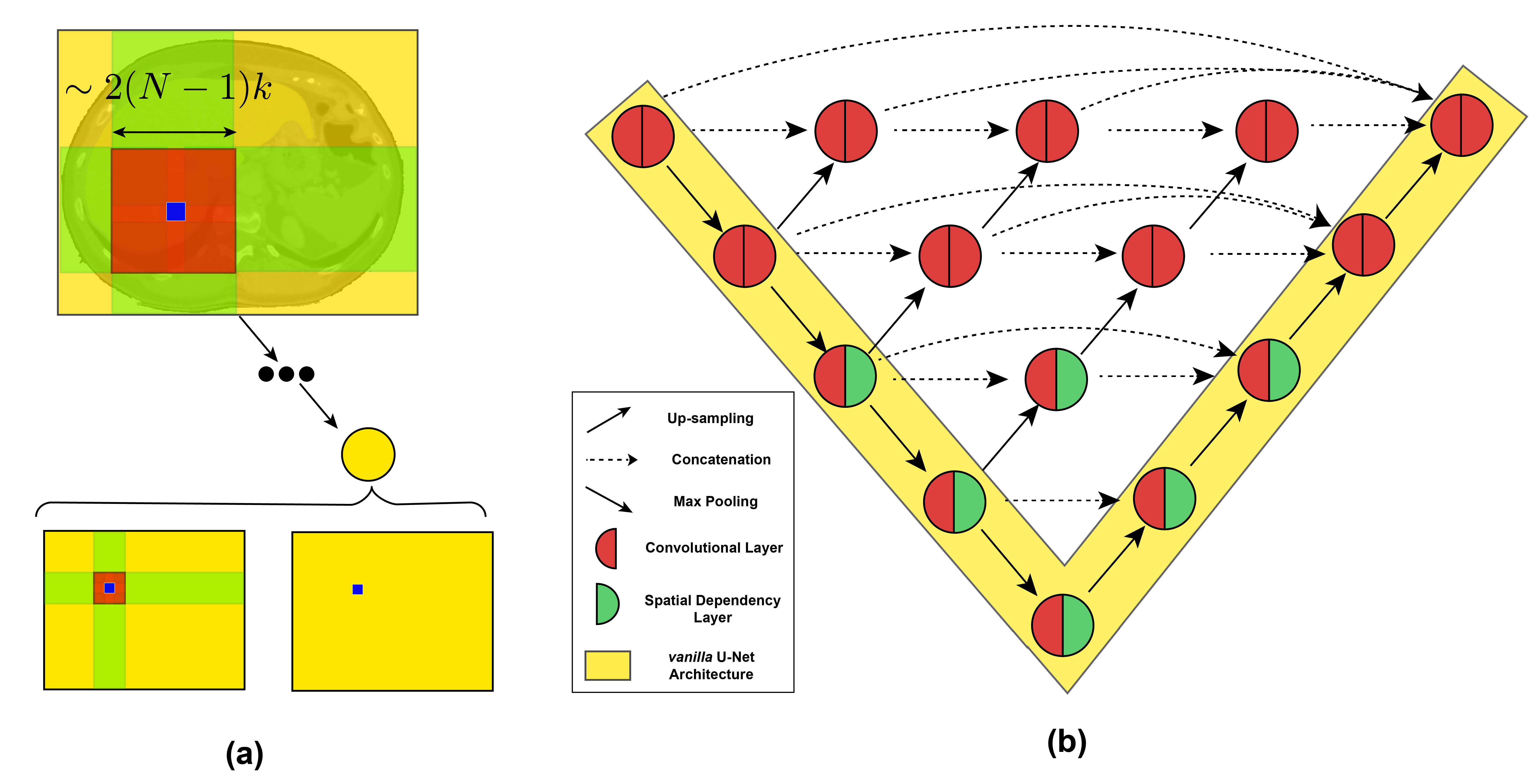}
\caption{\textbf{(a) Receptive field.} The receptive field of a conventional convolutional layer with kernel size $k\times k$ \textbf{(red)}, is $\sim2(N-1)k$ ($N$=number of layers). Due to the relative shallowness of medical segmentation architectures, it only partially encloses the original image, whereas the receptive field of spatial dependency layers \textbf{(green)} largely extends it bidirectionally. \textbf{(b) Diagram of architectures.} SDNU-Net integrates a mix of convolutional and spatial dependency layers into the deeper scales of U-Net++. In yellow we depict SDU-Net.} \label{figsdnu-net}
\end{figure}

\vspace{-0.3cm}
One of the major downsides in CNN based architectures is that direct interaction between features is limited to the convolutional kernel's size. Even though the degree of indirect interactions in deeper CNN layers is increased through pooling layers, these are still dependencies encompassing different semantic levels. On the other hand, spatial dependency layers' receptive field are not limited by kernel size: as sweeps are performed directionally across the full length of the image, feature dependencies are larger both within a single layer, and across multiple layers (as depicted in Fig. \ref{figsdnu-net} a)). 
We start by introducing the SDN's building blocks, and then connect them to our SDN based architectures. 

\subsection{Spatial Dependency Layers}
\begin{figure}
\includegraphics[width=\textwidth]{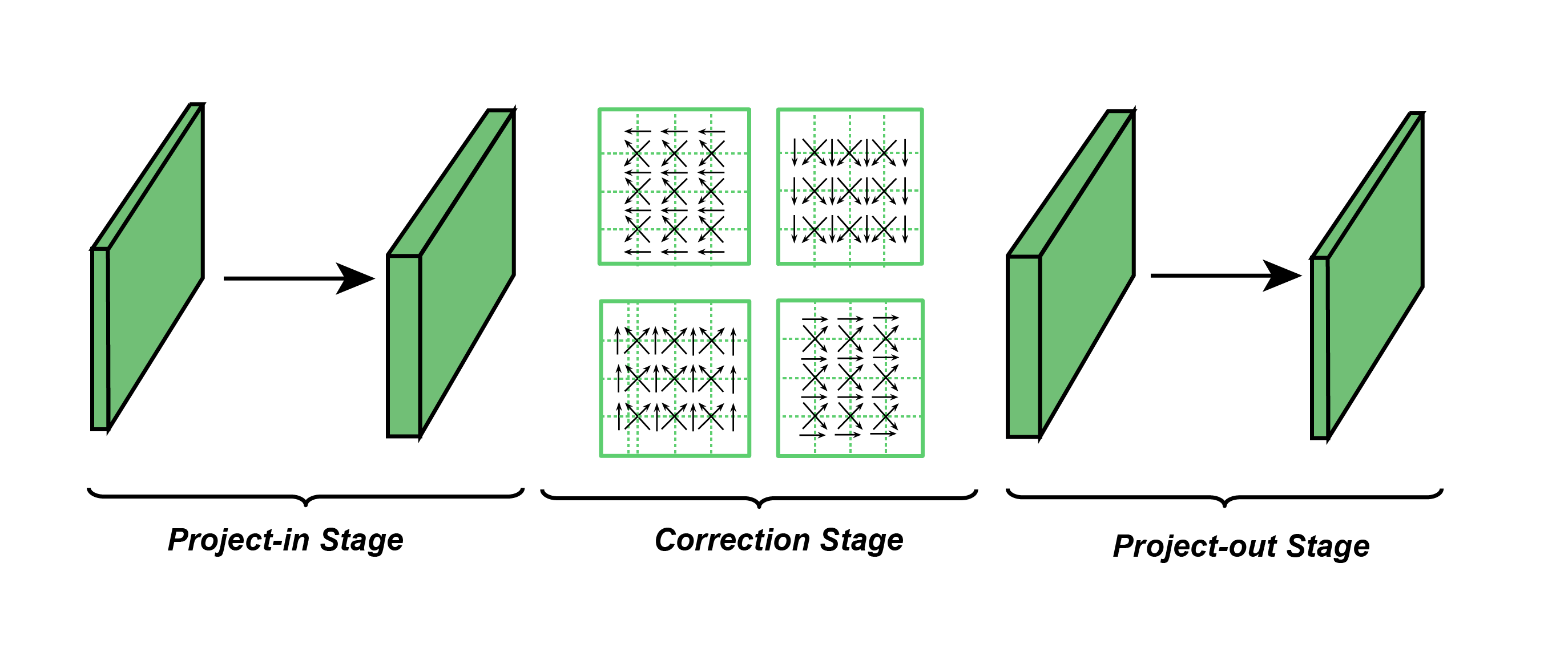}
\caption{\textbf{Three stages of spatial dependency layers}. Project stages are akin to convolutions with $1\times1$ kernels. The correction stage directionaly sweeps the feature map, updating the features while accounting for prior values. Figure adapted from \cite{miladinovic2021spatial}.} \label{figsdn_layer}
\end{figure}

The design of spatial dependency layers models a differential function that improves the modeling of spatial coherence and  high-order statistical correlations between pixels/voxels. Technically speaking, spatial dependency layers follow three steps: (1) \textit{project-in stage}, (2) \textit{correction stage}, and (3) \textit{project-out stage} (as shown in Fig. \ref{figsdn_layer}). 
Spatial dependency layers work independently of surrounding transformations, and similarly to convolutional layers can be integrated with both batch normalization and dropout. In the case of vanishing gradients during training, similarly to CNNs \cite{he2016deep}, an exchange with a variant equipped with a residual connection is possible \cite{miladinovic2021spatial}. 

\vspace{-.5cm}
\subsubsection{Projection Stages} Both projection stages implement a simple, vector-wise, affine transformation from an input feature map, $X^{s}$, to an output feature map, $X^{s+1}$:
\begin{equation}
    X^{s+1}_{i,j} = X^{s}_{i,j}\textbf{W} + \textbf{b},
\end{equation}
where $i$ and $j$ are the corresponding 2-D coordinates in $X^{s}$ and $X^{s+1}$, $\textbf{W}$ is a learnable weight matrix, and $\textbf{b}$ is a vector of learnable biases. Main distinctions from the \textit{project-in} and \textit{project-out} stages concern the number of channels in the output space: \textit{project-in stage} increases the number of channels from the input representation to a tunable number of channels, whereas \textit{project-out stage} maps features produced by the correction stage to an output feature map with a number of channels equal to the original input representation.

\subsubsection{Correction Stage} The goal of the correction stage is to update the feature values taking the intermediate representation produced by the \textit{project-in stage}, and sweeping the feature map directionally through recurrent units, where the channels can be interpreted as the state size. Four directions are considered in this work: left-right, right-left, up-down, and down up; these could also be extended to allow for the 6 directions consistent with 3D images. The sweeps are implemented using a gating mechanism \cite{cho2014gru} adapted to the image setting, which moderates the contributions of the updated \textit{(proposed)} value and the intermediate \textit{(prior)} feature value. Implementation details can be found in \cite{miladinovic2021spatial}.

\subsection{Introducing Spatial Dependency Layers to U-Net and U-Net++} 
Our main contribution to U-Net and U-Net++ is the integration of spatial dependency layers into both network architectures, while leveraging the benefits of both spatial dependency and convolutional layers by retaining the established design of U-Net and U-Net++ (see Fig.\ref{figsdnu-net} b)). U-Net++, by the incorporating nested dense connections to the U=Net architecture, mainly attempts to alleviate the drawbacks of features from different semantic levels being combined at the decoding path. This is done by redesigning the original architecture as an ensemble of U-Nets with different levels of depths, and concatenating feature maps from different scales through a redesigned dense skip-connection that aggregates feature maps from different semantic levels \cite{zhou2019unet++}. In our experiments, we found sufficient to include spatial dependency layers at lower scales of U-Net++, retaining most of the original nested architecture, and mitigating the computational complexity inherent to SDNs \cite{miladinovic2021spatial}. This was also verified for the vanilla U-Net. Our implementation consecutively applies convolutional and spatial dependency layers within the same depth level, leveraging both local and non-local similarity comparisons between feature patches at the same scale.

\section{Experiments}
\vspace{-0.5cm}
\begin{table}[ht]
\caption{Summary information of datasets used.} 
\label{tab:datasets}
\begin{center}
\begin{tabular}{lcccc}
                            & \textbf{\begin{tabular}[c]{@{}c@{}}Number of \\ Images\end{tabular}} & \textbf{\begin{tabular}[c]{@{}c@{}}Image Size \\ (resampled size)\end{tabular}} & \textbf{Modality} & \textbf{Challenge} \\ \cline{2-5} 
\multicolumn{1}{l|}{\textbf{Nuclei}} & \begin{tabular}[c]{@{}c@{}}670 \\ (2D images)\end{tabular}           & 96 $\times$ 96                                                                         & Microscopy        & 2018 Data Science Bowl \cite{caicedo2019nucleus}                                                  \\ \hline
\multicolumn{1}{l|}{\textbf{Polyps}} & \begin{tabular}[c]{@{}c@{}}612 \\ (29 sequences)\end{tabular}        & \begin{tabular}[c]{@{}c@{}}384 $\times$ 288 \\ (192 $\times$ 144)\end{tabular}                & Colonoscopy       & \begin{tabular}[c]{@{}c@{}}Endoscopic Vision\\ MICCAI 2015 \cite{bernal2017comparative}\end{tabular}  \\ \hline
\multicolumn{1}{l|}{\textbf{Liver}}  & \begin{tabular}[c]{@{}c@{}}131 \\ (3D volumes)\end{tabular}          & \begin{tabular}[c]{@{}c@{}}512 $\times$ 512 \\ (128 $\times$ 128)\end{tabular}                & CT                & \begin{tabular}[c]{@{}c@{}}LiTS\\ ISBI 2016/MICCAI 2017 \cite{lits_dataset}\end{tabular} 
\end{tabular}
\end{center}
\end{table}
\vspace{-1cm}
\subsection{Datasets}
As described in Table \ref{tab:datasets}, three public medical image segmentation datasets were used to evaluate the proposed architectures, which cover different medical imaging modalities and segmentation tasks. Partitioning between train, validation, and test set was always done at the volume/patient level to avoid bias and over-optimistic result. For further details on the datasets and corresponding data pre-processing, please refer to the supplementary material.

\subsection{Implementation details}
SDU-Net and SDNU-Net were benchmarked in these three segmentation tasks against the widely used U-Net and U-Net++. All models were trained with a combination of dice and cross-entropy as its loss function \cite{isensee2018no}. Both Dice and Jaccard score were monitored during training, with the final model being selected through \textit{early-stopping} on the validation sets. All experiments were done in a computational configuration consisting of four Nvidia RTX 2080 Ti GPUs (11GB/GPU). Following common practices from the field \cite{zhou2019unet++,isensee2018no}, the number of layers of each network architecture was tuned to each segmentation task. Similarly, SD specific parameters - state size, \textit{i.e.} the number of channels in each \textit{project-in stage} of spatial dependency layer, number of directions, and overall number of layers equipped with spatial dependency were optimized. Architecture details of baseline models, including activation functions and kernel sizes, followed original descriptions in \cite{zhou2019unet++,ronneberger2015unet}. Final configurations and implementation details are disclosed in the supplementary material.

\subsection{Results}
\vspace{-0.5cm}
\subsubsection{Comparison with baselines}

\begin{table}[]
\label{tab:results1}
\caption{Segmentation results averaged across 5 runs (Dice and Jaccard score: \%), for baseline models (U-Net and U-Net++) and SDN models (SDU-Net and SDNU-Net).}
\setlength{\tabcolsep}{2.5pt}
\renewcommand{\arraystretch}{1.7}
\begin{tabular}{lc c c c cc}
& \multicolumn{2}{c}{\textbf{Nuclei}} & \multicolumn{2}{c}{\textbf{Polyps}} & \multicolumn{2}{c}{\textbf{Liver}}                    \\ \cline{2-7} 
\multicolumn{1}{l}{\textbf{Model}}    & \textbf{Dice}   & \textbf{Jaccard}   & \multicolumn{1}{|c }{\textbf{Dice}}   & \textbf{Jaccard}     &\multicolumn{1}{|c }{\textbf{Dice}} & \textbf{Jaccard} \\ \hline
\multicolumn{1}{l|}{U-Net}    & 91.79$\pm$0.32    & \multicolumn{1}{l|}{85.52$\pm$0.50}  & 75.37$\pm$0.60    &\multicolumn{1}{c|}{65.32$\pm$0.67}       & 77.44$\pm$2.25 & 69.92$\pm$2.19     \\ 
\multicolumn{1}{l|}{U-Net++}  & 92.64$\pm$0.24    & \multicolumn{1}{l|}{86.88$\pm$0.38}  & 76.33$\pm$1.69    &\multicolumn{1}{c|}{66.45$\pm$1.73}       & 80.24$\pm$2.10 & 73.73$\pm$2.46     \\ 
\multicolumn{1}{l|}{SDU-Net}  & 93.25$\pm$0.35    & \multicolumn{1}{l|}{87.33$\pm$0.58}  & 80.62$\pm$0.79    &\multicolumn{1}{l|}{71.04$\pm$0.90}       & 82.43$\pm$2.24 & 75.02$\pm$2.51     \\ 
\multicolumn{1}{l|}{SDNU-Net} &\underline{94.10}$\pm$\underline{0.36}    & \multicolumn{1}{l|}{\underline{88.71}$\pm$\underline{0.52}}  & \underline{83.31}$\pm$\underline{1.30}    &\multicolumn{1}{l|}{\underline{73.19}$\pm$\underline{1.55}}       & \underline{85.72}$\pm$\underline{2.45} & \underline{79.37}$\pm$\underline{2.17}  \\ \hline
\end{tabular}
\end{table}
Table \ref{tab:results1} compares baselines, U-Net and U-Net++, with proposed models, SDU-Net and SDNU-Net, in terms of segmentation performance measured in Dice and Jaccard score (mean$\pm$ s.d.) over the three segmentation tasks. Stability of the model performance was assessed through averaging results over five runs (training, model selection through early-stopping, and evaluation in test-set).

Globally, the inclusion of spatial dependency layers improves model performance in all evaluated segmentation tasks, with both SDU-Net and SDNU-Net obtaining enhanced performance comparatively to its baselines (average improvement of 4.29 Dice points and 4.47 Jaccard points). The vanilla U-Net is also outperformed in most metrics by our U-Net equipped with spatial dependency layers (SDU-Net). In both the colon polyp segmentation and the liver segmentation setting large improvements were shown between SDNU-Net and U-Net++, with a 6.98 Dice and 6.74 Jaccard gain for the polyp segmentation, and 5.48 Dice and 5.64 Jaccard increase for the liver segmentation. Arguably this may be due to the variable scales at which polyps and liver cross-sections appear, with large texture changes between both polyps and 'normal tissue', i.e. non-tumoral, and liver and non-liver tissue. 
Qualitative comparison in Fig. \ref{figqualitative_comparison} also corroborates this assertion.

\subsubsection{Ablation}
\begin{figure}
\centering
\includegraphics[width=0.8\textwidth]{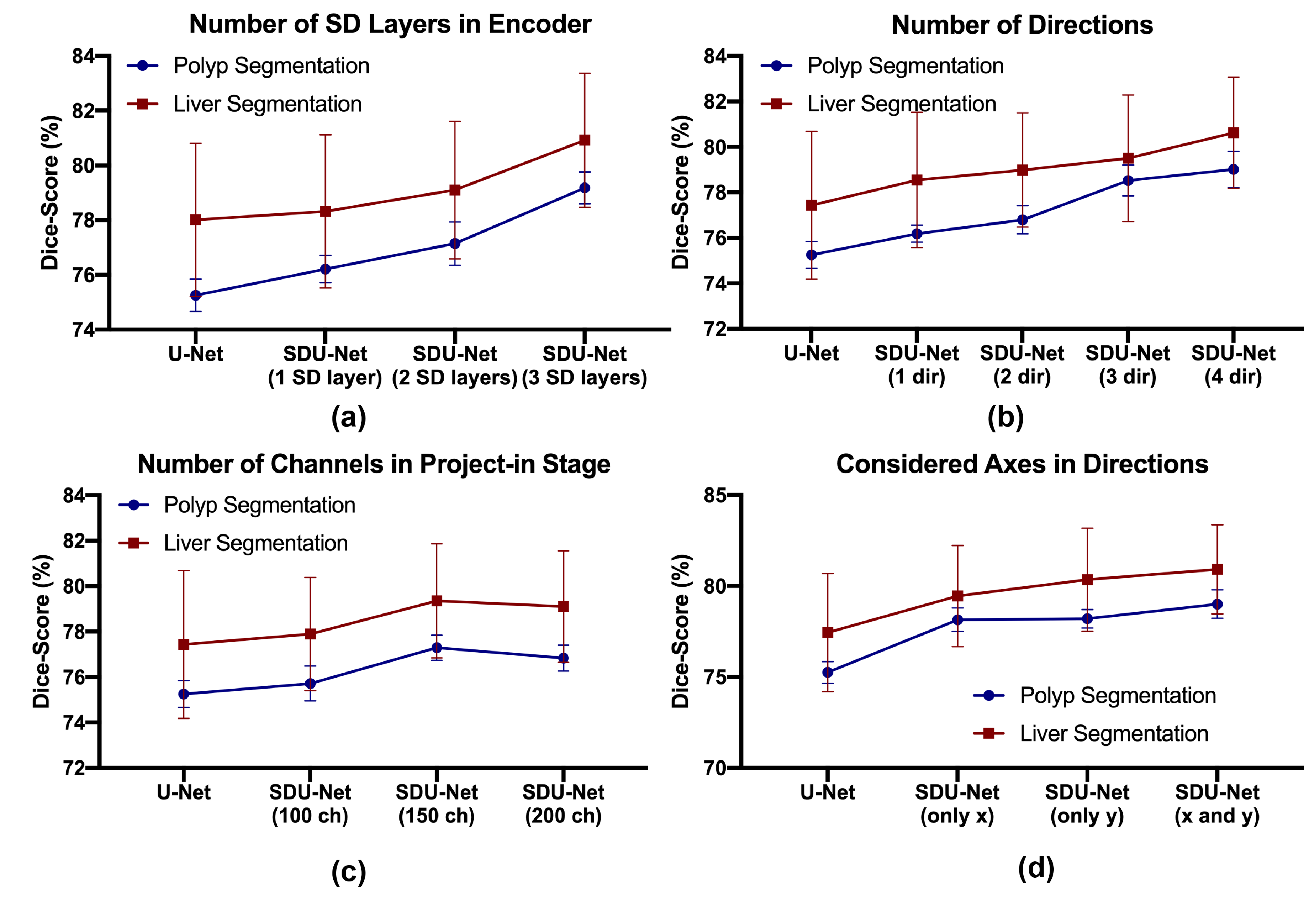}
\caption{\textbf{Ablation studies.} Fixing SDU-Net with 1 spatial dependency layer in the encoder, direction left-to-right, and 100 output channels in the \textit{project-in stage}, all hyperparameters were scanned, and models compared to a 4 layer U-Net.} \label{figablation}
\end{figure}
Contributions of hyper-parameter tuning were assessed through a series of ablations studies as follows. Fixing the general architecture design, and evaluating in the colon polyp segmentation and liver segmentation tasks, contributions from SD specific parameters were estimated. We verify that increased number of scales equipped with a spatial dependency layers (Fig.\ref{figablation} a)), number of directions (Fig.\ref{figablation} b)), and number of output channels in the project-in stage (Fig.\ref{figablation} c)) lead to overall improvement in performance. Results additionally suggest that the choice of sweeping directions (Fig.\ref{figablation} d)) also impacts performance. In specific, for both segmentation tasks performing two sweeps across two different axes is preferential to sweeping bidirectionally across the same axis. 

\section{Conclusion}
In this work we propose two novel SDN based architectures (SDNU-Net and SDU-Net) that take advantage on increased spatial coherency in feature maps and better modeling of long distance dependencies between features, both critical for medical imaging segmentation tasks. Experiments in three segmentation tasks: nuclei segmentation in microscopy images, colon polyp segmentation in colonoscopy videos, and liver segmentation in abdominal CT scans demonstrate superior performance of both models, with an average increase of 4.29 Dice points and 4.47 Jaccard points across all tasks, and even higher improvement for the polyp and liver segmentation tasks. Simple integration of spatial dependency layers points to our Spatially Dependent U-Nets being broadly generalizable architectures in the medical imaging segmentation domain.

% ---- Bibliography ----
%
% BibTeX users should specify bibliography style 'splncs04'.
% References will then be sorted and formatted in the correct style.
%
% \bibliographystyle{splncs04}
\bibliography{bibliography}

\newpage

\section{Supplementary Material}
\begin{table}[]
\caption{Relevant dataset statistics and main pre-processing steps.}
\renewcommand{\arraystretch}{1.7}
\setlength{\tabcolsep}{7pt}
\setstackEOL{\#}% LINE SEPARATOR IN STACK (\\ CONFLICTS WITH TABULAR USE)
\renewcommand\stacktype{L}% LONG STACKS ARE THE DEFAULT
\setstackgap{L}{11pt}% THIS IS THE VERTICAL BASELINESKIP IN LONG STACKS
\begin{tabular}{lccc}
\textbf{\begin{tabular}[c]{@{}l@{}}Segmentation Task\end{tabular}} & \textbf{nuclei}                                                                                    & \textbf{polyps}                                                               & \textbf{liver}                                                               \\ \hline
\multicolumn{1}{l|}{Dataset name}                                     & DSB2018                                                                                            & CVC-ClinicDB                                                                   & LiTS                                                                         \\
\multicolumn{1}{l|}{\multirow{2}{*}{Image type}}                      & \multirow{2}{*}{\Longunderstack[c]{brightfield and \# fluorescent microscopy}} & \multirow{2}{*}{\Longunderstack[c]{colonoscopy \# video}} & \multirow{2}{*}{\Longunderstack[c]{abdominal \# CT scan}} \\
\multicolumn{1}{l|}{}                                                 &                                                                                                    &                                                                               &                                                                              \\
\multicolumn{1}{l|}{Dataset size}                                     & 670 images                                                                                         & \Longunderstack[c]{29 sequences \# (612 frames)}           & 131 volumes                                                                  \\
\multicolumn{1}{l|}{Original size}                                    & 96 x 96                                                                                                 & 304 x 288                                                                     & 512/512                                                                      \\
\multicolumn{1}{l|}{Resampled size}                                   & -                                                                                                  & 192 x 144                                                                     & 128 x 128                                                                    \\
\multicolumn{1}{l|}{Data split$^{1}$}                                       & \Longunderstack[c]{80/10/10 \# (\%)}                                            & \Longunderstack[c]{23/3/3 \# (sequences)}                 & \Longunderstack[c]{82/21/28 $^{2}$  \# (volumes)}                \\
\multicolumn{1}{l|}{Intensity clipping}                               & -                                                                                                  & -                                                                             & {[}-1000; 1000{]} HU                                                         \\
\multicolumn{1}{l|}{Rescale to {[}0,1{]}}                             & Yes                                                                                                & Yes                                                                           & Yes     \\ \hline         
\vspace{-0.5cm} \\ \cline{1-2}
\multicolumn{4}{l}{$^{1}$ \footnotesize{partition at sequence level for CVC-ClinicDB and at the volume/patient level for LiTS. }}\\

\multicolumn{4}{l}{$^{2}$ \footnotesize{from the two original available training batches, one (103 volumes) was split 80/20 into}}\\
\vspace{-1.0cm}\\
\multicolumn{4}{l}{\footnotesize{training/validation, and the other (28 volumes) was used for testing.}}\\
\end{tabular}
\end{table}

\begin{table}[hb]
\caption[Compact Routing Example]%
{Comparison of model size, in number of parameters, and inference computation cost, in multiply-accumulate-operation (MAC), for all models evaluated.}
\renewcommand{\arraystretch}{1.5}
\setlength{\tabcolsep}{4pt}

\begin{tabular}{lcccc}

                                  & \textbf{U-Net} & \textbf{U-Net++} & \textbf{SDU-Net} & \textbf{SDNU-Net} \\ \hline
\multicolumn{1}{l|}{\textbf{Number of parameters (M)}}    &                &                  &                  &                   \\
\multicolumn{1}{l|}{nuclei}                               & 1,8            & 2,24             & 5,42             & 6,16              \\
\multicolumn{1}{l|}{polyps}                               & 7,24           & 9,76             & 10,36            & 11                \\
\multicolumn{1}{l|}{liver}                                & 7,24           & 9,76             & 10,36            & 11                \\ \hline
\multicolumn{1}{l|}{\textbf{Number of operations (Gmac)$^{3}$}} &                &                  &                  &                   \\
\multicolumn{1}{l|}{nuclei}                               & 1,31           & 2,87             & 2,59             & 4,7               \\
\multicolumn{1}{l|}{polyps}                               & 2,28           & 6,5              & 3,01             & 8,15              \\
\multicolumn{1}{l|}{liver}                                & 1,71           & 4,87             & 2,36             & 6,09              \\ \hline
\vspace{-0.4cm} \\ \cline{1-2}
\multicolumn{5}{l}{$^{3}$ \footnotesize{computed using \textit{ptflops}, (available at: github.com/sovrasov/flops-counter.pytorch)}}
\end{tabular}

\end{table}

\begin{table}[]
\caption{Experimental configurations of SDU-Net and SDNU-Net.}
\renewcommand{\arraystretch}{1.5}

\begin{tabular}{lcccccc}

\textbf{}                                                     & \multicolumn{3}{c}{\textbf{SDU-Net}}                & \multicolumn{3}{c}{\textbf{SDNU-Net}} \\ \hline
\multicolumn{1}{l|}{\textbf{Segmentation task}}               & \textbf{nuclei}   & \textbf{polyp}   & \multicolumn{1}{c|}{\textbf{liver}}    & \textbf{nuclei}      & \textbf{polyp}     & \textbf{liver}      \\
\multicolumn{1}{l|}{Optimizer}                       & Adam     & Adam     & \multicolumn{1}{c|}{Adam}     & Adam        & Adam       & Adam       \\
\multicolumn{1}{l|}{Learning rate}                   & 1,00E-03 & 1,00E-04 & \multicolumn{1}{c|}{1,00E-04} & 1,00E-03    & 1,00E-04   & 1,00E-04   \\
\multicolumn{1}{l|}{Convolutional layer kernel size} & 3x3      & 3x3      & \multicolumn{1}{c|}{3x3}      & 3x3         & 3x3        & 3x3        \\
\multicolumn{1}{l|}{Activation function}             & ReLU     & ReLU     & \multicolumn{1}{c|}{ReLU}     & ReLU        & ReLU       & ReLU       \\
\multicolumn{1}{l|}{Batch Normalization}             & Yes      & Yes      & \multicolumn{1}{c|}{Yes}      & Yes         & Yes        & Yes        \\
\multicolumn{1}{l|}{Number of layers per scale}      & 2        & 2        & \multicolumn{1}{c|}{2}        & 2           & 2          & 2          \\
\multicolumn{1}{l|}{Residual Connections}            & No       & Yes      & \multicolumn{1}{c|}{Yes}      & No          & Yes        & Yes        \\
\multicolumn{1}{l|}{SDN kernel size}                 & 3x3      & 3x3      & \multicolumn{1}{c|}{3x3}      & 3x3         & 3x3        & 3x3        \\
\multicolumn{1}{l|}{SDN \#channels per scale}        & 150      & 150      & \multicolumn{1}{c|}{150}      & 150         & 150        & 150        \\
\multicolumn{1}{l|}{SDN layers per encoder/decoder}  & 2        & 1        & \multicolumn{1}{c|}{1}        & 2           & 1          & 1          \\
\multicolumn{1}{l|}{Number of directions per SDN}    & 2        & 2        & \multicolumn{1}{c|}{2}        & 2           & 2          & 2          \\ 
\multicolumn{1}{l|}{Batch sizer per GPU}    & 20        & 20        & \multicolumn{1}{c|}{16}        & 20           & 20          & 16          \\ 
\multicolumn{1}{l|}{Number of GPUs}    & 4        & 4        & \multicolumn{1}{c|}{4}        & 4           & 4          & 4          \\ 
\multicolumn{1}{l|}{GPU memory (GB)}    & 11        & 11        & \multicolumn{1}{c|}{11}        & 11           & 11          & 11          \\  \hline

\end{tabular}
\end{table}

\end{document}